\documentstyle[preprint,aps]{revtex}

\begin{document}

\preprint{
\font\fortssbx=cmssbx10 scaled \magstep2
\hbox to \hsize{
\hbox{\fortssbx University of Wisconsin - Madison}
\hfill$\vcenter{\hbox{\bf MADPH-97-1013}
                 \hbox{\bf UH-511-882-97}
                \hbox{September 1997}}$ }
}

\title{\vskip.5in
Are $e\mu$ colliders interesting?}
\author{V. Barger$^1$, S.~Pakvasa$^2$, and X.~Tata$^2$}
\address{$^1$Physics Department, University of Wisconsin, Madison, WI 53706\\
$^2$Department of Physics and Astronomy, University of Hawaii, Honolulu,
HI 96822}

\maketitle

\begin{abstract}
We show that current experimental constraints already severely restrict what
might be observable at $e\mu$ colliders. We identify some cases where it may be
possible to probe physics beyond what might be possible at other facilities and
make some remarks about physics capability of high energy $e\mu$ colliders.
\end{abstract}

\thispagestyle{empty}

\newpage

Recently, Hou\cite{hou} and Choi et al.\cite{choi} have suggested the
intriguing possibility of searching for lepton flavor violating (LFV) couplings
at $e\mu$ colliders. They propose to search for $e$ and $\mu$ number violating
couplings of hypothetical heavy particles $X$ that may be produced as
resonances in $e\mu$ collisions. The implementation of this idea is hampered by
our complete ignorance of the mass $M_X$, {\it i.e.}\ of not knowing the energy
at
which to operate the collider.

In this note, we first explore the more practical idea of looking for LFV
interactions by searching for resonance production of {\em known} particles,
which has the obvious advantage that we know the exact center of mass (CM)
energies at which the collider should be operated.  The other very important
advantage of this idea is that we would {\em not} require a very high energy
muon beam since most of the known resonances are lighter than about 10~GeV. We
were especially motivated to examine this since it may well be that the
development of cold high intensity muon beams with $E\sim 1$~GeV could be a
needed first step for the development of a high energy muon collider.
Unfortunately, we find that current limits on LFV interactions of known
particles already put severe limits on the cross sections for producing these
via $e\mu$ collisions, so that with the preliminary estimates\cite{palmer}
for the luminosity
\begin{equation}
{\cal L} \sim 2\times10^{32} \left( E\over 100{\rm\ GeV} \right)^{4/3} \rm
cm^{-2} \, s^{-1}   \label{lum}
\end{equation}
the expected rates, with 1--10~GeV $e\mu$ colliding beams, are generally
well below the
level of observability. Following Choi et al.\cite{choi} we next consider high
energy $e\mu$ colliders designed to operate at the resonance $X$. We show that
rates for LFV processes will be strongly limited if $X$ also  couples to
hadrons. For completeness, we point out exceptional scenarios where there could
be observable rates at $e\mu$ colliders, but which cannot be probed at high
energy $e^+e^-$ (and even $\mu^+\mu^-$) or hadron colliders.

\bigskip\leftline{\bf Resonance Production of Known Particles}

The best limits on $e\mu$ flavor violation came from the non-observation
of the reaction $\mu Ti \rightarrow e Ti$, or the decays
$\mu\to e\gamma$ and $\mu\to 3e$. The branching fraction for
these decays are smaller\cite{barnett} than $5\times10^{-11}$ and $10^{-12}$,
respectively.  These bounds strongly limit the LFV $e\mu$ couplings of
flavor-neutral vector mesons such as $\rho$, $\psi$, $\Upsilon$ (and
their excitations) which directly couple to single photons. It is easy
to check, for example, that the limit on $B(\mu\to e\gamma)$ leads to
the remarkably strong bound ${f_{\rho\mu e}^2\over 4\pi} < 10^{-26}$, to
be compared with ${f_{\rho\pi\pi}^2\over 4\pi} < 2.5$.

Spin-zero mesons cannot directly couple to the photon but can couple to
$e^+e^-$ pairs, and lead to $\mu\to 3e$ decays if these mesons have LFV $\mu e$
interactions. This decay cannot proceed via 1-photon exchange (even for the
$0^+$ state) since the electromagnetic current is exactly conserved, but occurs
via multi-photon exchange or via the $Z^0$ exchange, and is further suppressed
by the chirality factor $(m_e/m_X)^2$. The best limit on LFV interactions of
spin-zero mesons appears to come from the non-observation of the decay $\mu\to
e\gamma\gamma$, the branching fraction for which is
smaller\cite{barnett}
than $7\times
10^{-11}$. For example, for $X=\eta$, using $\Gamma(\eta\to\gamma\gamma)\simeq
0.5$~keV, we obtain ${f_{X\mu e}^2\over 4\pi} \alt 10^{-17}$. A similar bound
should apply on LFV couplings of other spin-zero mesons, assuming only that
$f_{X\gamma\gamma} \simeq f_{\eta\gamma\gamma}$.

There are direct limits\cite{barnett} on $\mu e$ LFV decays of flavored
mesons $K_L^0$, $D^0$ and $B^0$. The strongest of these limits is on the
branching fraction for $K_L^0\to \mu e$ decay which is smaller than
$3\times10^{11}$. The corresponding bounds for $\mu e$ decay of $D^0$
($B^0$) mesons are $\sim\rm few \times 10^{-5}\ (few\times 10^{-6})$. It
is straightforward to check that the direct limit on $B(K_L^0\to\mu^\pm
e^\mp)$ is much more restrictive than the limit that would be obtained
from $\mu\to e\gamma\gamma$ decay using the observed decay rate for
$K_L^0\to\gamma\gamma$.

The decays $K_L\to\mu^\pm e^\mp$ probe different sources of LFV than do the
corresponding decays of $K^*$ on which the bounds are considerably weaker.
Nevertheless, from the limit\cite{barnett} $B(K^+\to\pi^+\mu^+e^-)\alt
2 \times 10^{-10}$ we may infer bounds such as $B(K^*\to\mu^+ e^-)\alt
10^{-21}$, since
otherwise the rate for the decay $K^+\to\pi^+\mu^+e^-$ mediated by virtual
$K^*$ would exceed its experimental bound. It should be amply clear that a
similar bound applies to {\em all} strange resonance $K_X$ with $J^P = 0^+,
1^-, 2^+,\dots$ since the $KK_X\pi$ vertex is allowed by strong interactions.
We thus conclude that the $K_X\mu e$ coupling can only be significant for
strange mesons with $J^P = 0^-, 1^+, 2^-$ {\it etc.} Within the quark model
framework, however, this possibility also appears to be excluded since any
quark bilinear $\bar s\Gamma d$ with non-vanishing matrix elements between
these $J^P = 0^-, 1^+, 2^-$ etc.\ states and the vacuum would also result in
$K_L^0\to\mu e$ decays, unless various contributions cancel to a very high
precision, or form factors become tiny for no apparent reason. We thus
conclude that $\mu e$ LFV couplings of strange mesons with all $J^P$ quantum
numbers consistent with the quark model are very small.

We now turn to the examination of what might be possible at $e\mu$ colliders.
The peak cross section for resonance production of a particle $X^0$ with spin
$S=(0$ or 1) is given by
\begin{equation}
\sigma = {4\pi\over M_X^2} (2S+1) B_{e\mu} \,,  \label{sigmaX}
\end{equation}
where $B_{e\mu}$ is the branching fraction for the decay $X^0\to
e^-\mu^+$ (or $e^+\mu^-$), depending on the initial
beams. Eq.~(\ref{sigmaX}) presumes that the spread in energy ($\Delta$)
is much smaller than the width $\Gamma$ of the resonance. In the case that
$\Delta \gg \Gamma$, the effective luminosity, and hence the event rate at the
peak, is reduced by a factor $\sim\Gamma/\Delta$, the exact number
depending on the beam profile. In what follows, we take $\Delta \sim
10$~MeV, which for $\sqrt s = 1$--10~GeV corresponds to a beam
resolution of (0.1--1)\%, to be compared with the the
projected\cite{proposal} beam resolution of better than 0.1\% for muon
beams and typical resolutions of a $\rm few\times10^{-4}$ at existing
$e^+e^-$ colliders. In order to maximize the event rate, it is clear
that we should focus on particles with widths larger than 10~MeV.

The most obvious $X$ candidate is $Z^0$ which, however, is excluded for reasons
that we have already mentioned in another context: the experimental bound
$B(\mu\to 3e) < 10^{-12}$, in turn, limits $B(Z^0\to\mu^+e^- + e^-\mu^+) <
6\times 10^{-13}$. From Eq.~(\ref{sigmaX}) it is then straightforward to check
that even with an integrated luminosity of 100~fb$^{-1}$, we would expect $\alt
0.1$ event at an $e\mu$ collider operating on $Z^0$.

We are thus led to examine the possibility of producing known hadrons in $\mu
e$ collisions. For a CM energy $E\sim 1$~GeV, Eq.~(\ref{lum}) yields an
integrated luminosity of $\sim4\rm~pb^{-1}$, assuming collider operation for
$10^7$\,s. Using Eq.~(\ref{sigmaX}), we see that for resonances with $\Gamma
\agt \Delta$, we may expect about
\begin{equation}
{\cal N} \simeq (2S+1)  B_{e\mu} (2\times 10^{10}) / M^{2/3} \label{N}
\end{equation}
events (here $M$ is in GeV units) during this period of operation, so
that
we can at best probe LFV decays with a branching fraction
$\alt10^{-10}$ for $M\sim1$--5~GeV. But our previous discussion shows that
current experimental bounds already essentially exclude this range for most
known resonances. LFV decays of strongly decaying flavor-neutral mesons, we
saw, were constrained to have branching fractions $\alt10^{-26}$, while the
bounds on their pseudoscalar counterparts were $\sim10^{-17}$. LFV couplings of
$1^+$ scalar bosons would have similar bounds on their couplings as $0^-$
bosons. We have also seen that the limits on LFV decays $K_L^0\to\mu e$ and
$K_L^0\to\pi\mu e$ respectively limit the LFV branching fractions of $0^-, 1^+,
2^-\dots$  ($0^+, 1^-, 2^+\dots$) mesons to be smaller than $3\times10^{-11}$
($\sim10^{-21}$), which again would be below the level of observability given
in Eq.~(\ref{N}), unless luminosities significantly higher than those given by
(\ref{lum}) are achieved.

Despite the fact that we have arrived at a generally negative assessment
regarding the feasibility of observing known resonances in $e\mu$ collisions,
there is one loophole in our arguments up to now. Recall that LFV interactions
of unflavored $1^+$ mesons were constrained only by upper limits on the $e\mu$
decay rate of the corresponding pseudoscalar state. There are, however, no such
limits on $\eta_c$ and $\eta_b$ decays. This leads us to suggest that at $e\mu$
colliders it may be possible to probe LFV couplings of $\chi_{c1}$ ($J^{PC} =
1^{++}$) which has a width $\sim0.9$~MeV. Taking into account the suppression
from the factor $\Gamma/\Delta$, we see from (\ref{N}) that optimistically it
should be possible to probe $B(\chi_{c1}\to e\mu)$ down to about $\rm few
\times 10^{-10}$ since there is no physics background to the signal
$e\mu\to\chi_{c1}\to\rm anything$, where the invariant mass of the final state
reconstructs to $M(\chi_{c1})$. Electron contamination from the decays of muons
in the beam would lead to hadronic signals with smaller invariant mass. A
$\sim4\pi$ detector would thus be necessary to convincingly study the signal.
We have checked that the current bounds on $B(\pi^0\to\mu e)$ do not constrain
the hypothetical $\chi_{c1}\to \mu e$ couplings. We note that $\chi_{b1}$
states are expected to be somewhat narrower so that the range of branching
fractions that may be probed via $e\mu$ collisions is smaller by a factor of
5--10.

Finally, we turn to bare charm and bottom mesons. Current
limits\cite{barnett}
on the decays
$B^0\to\mu e$, $D^0\to\mu e$ are in the vicinity of $\rm few\times
[10^{-6}\mbox{--}10^{-5}]$, while limits on $B^+$ or $D^+$ decays to $\pi e\mu$
are $\sim\rm few\times 10^{-3}$. As we discussed for the kaon system, any
significant LFV $\mu e$ couplings of $0^+, 1^-, 2^+\dots$ states of this system
results in branching fractions for $D\to\pi e\mu$ decays in excess of
experimental bounds, just because the $D$ states are so narrow. The same
considerations hold for $B$ mesons. We thus focus on LFV couplings of $0^-,
1^+, 2^-\dots$ $D$ and $B$ mesons, which are small enough to escape the direct
bounds, and whose widths are larger than $\sim10$~MeV. The only established
state that we could find was $D_1(2420)$ $[J^P=1^+]$, whose width is 19~MeV.
Using (\ref{N}), we see that it should be possible to probe a branching
fraction of $\sim10^{-10}$ after a year of $e\mu$ collider operation, to be
compared with branching fraction limits of $\sim10^{-5}$ available today, and
at best of $\sim10^{-8}$ that may be possible\cite{LiuPak} at the Tevatron or
charm meson factories. There is no suitable $B$ meson state that has been
clearly identified, though it is quite possible that such a state may be
discovered at $B$-factories. We also note that in principle $\mu^+e^-$ and
$\mu^-e^+$ collisions probe LFV couplings that are a priori independent.

\bigskip\leftline{\bf {\boldmath\mbox{$e\mu$}} Colliders at High Energy}

Although this is not the main focus of this present study, we  make a few
remarks on $e\mu$ collider operation at very high energy where the elementary
carrier of LFV interactions can be resonantly produced. For an integrated
luminosity of 100~fb$^{-1}$, the number of events in the final state $f$
(assuming the collider is operated at the resonance peak) is given by
\begin{equation}
N = {0.5(2S+1) \times 10^9 B_{\mu e} B_f\over M_X (\rm TeV)^2} \,, \label{Nf}
\end{equation}
where $B_{\mu e}$ and $B_f$ are the branching fractions for the decays $X\to
\mu e$ and $X\to f$, respectively\cite{note}. We see from (\ref{Nf}) that the
event rate can be $\sim10^8/100\rm~fb^{-1}$ even for $M_X\sim$~TeV. This is not
necessarily in conflict with limits on the branching fraction for $\mu\to 3e$
decay, which is given by
\begin{equation}
B(\mu\to 3e) = K \left( g_{\mu e} g_{ee}\over M_X^2\right)^2 {1\over G_F^2} \,,
\end{equation}
where the constant $K$, which is ${\cal O}(1)$ depends on the spin and assumed
spacetime structure of the interactions of $X$ with electrons and muons. If
$g_{\mu e} \sim g_{ee}\alt 10^{-3}$, $B(\mu\to 3e)$ is quite compatible with
experimental bounds, {\em and} we can have a large event rate in (\ref{Nf}) if
$B_{\mu e} \simeq B_{\mu\mu}\simeq 1/2$. While couplings $\sim10^{-3}$ may
appear small for gauge interactions, it is worth keeping in mind that $X$ could
be a new spin-zero boson with modest couplings to leptons. Our message is that
the physics of LFV interactions, should these exist, is completely unknown.

We also remark that $e\mu$ LFV interactions of $X$,
are stringently constrained if $X$ also couples to hadrons.
In
this case $X$ will induce current-current LFV interactions, which as we saw in
the previous section, are strongly constrained by experiment.
For example, the vector
component of the hadronic current that couples to $X$ leads to $\mu\to e\gamma$
if it is flavor-neutral, or to $K\to\pi\mu e$ decays if it is $\bar d\gamma_\mu
s$, etc. Similarly, there should be strong bounds on LFV couplings of
sneutrinos if these also couple to hadrons in SUSY models where $R$-parity is
not conserved.

Finally, we remark on whether it would be possible to study the
resonance production of $X$ at other colliders. This, of course, depends
on what $X$ couples to. If it couples to $e^+e^-$ and $e\mu$ pairs, the
cross section for $e^+e^-\to X\to\mu e$ and $\mu e\to X\to e^+e^-$
should be equal. Since both processes are free from physics backgrounds
we see no particular advantage of $e\mu$ colliders. If instead $X$
couples to $\mu^+\mu^-$ and $e\mu$ pairs, it can be searched for at
$\mu^+\mu^-$ colliders. If, however, it only couples to $e\mu$ and
$\tau^+\tau^-$ pairs, $e\mu$ colliders appear to provide the only way of
directly searching for it. These colliders are also the unique
facility\cite{hou} to study the {\it direct} production of a
hypothetical particle responsible for muonium--antimuonium
oscillations\cite{FN}, as long as it does not also couple to other
channels.

What if $X$ couples to hadrons? One would then think that it would be simple to
search for it at the LHC via its $\mu e$ decays. Such a search would be akin to
the search for $Z'$ bosons except that the final state would be even more
characteristic. We note, however, that at the LHC it is possible\cite{atlas}
to search for a $Z'$ of mass 1~TeV only if $(\sigma\cdot B)_{Z'}
\agt(2\times10^{-3}) (\sigma\cdot B)_Z$, assuming an integrated luminosity of
100~fb$^{-1}$.
We thus warn the reader that if the couplings of $X$ to first generation
quarks $\alt\rm few\times 10^{-3}$ it could escape detection at the LHC, but,
depending on its decay patterns may be observable at an $e\mu$ collider.

\break\leftline{\bf Summary}

We have shown that existing constraints on LFV interactions severely restrict
what might be observable at $e\mu$ colliders. Our arguments leave some
loopholes for LFV interactions with special form of flavor and spacetime
structure. In these cases, a {\em low energy} $\mu e$ collider operating at the
resonance energy of $\chi_{c1}(3150)$, $\chi_{b1}(9892)$, or $D_1(2420)$ may be
able to probe LFV $\mu e$ couplings beyond current bounds, and likely beyond
what might be possible at other facilities. Moreover, our arguments do
not apply to mesons with quantum numbers that do not correspond to those
of quark-antiquark bound states as given by the quark model\cite{bnl}.
While we regard the theoretical
case for such LFV interactions as far from compelling, our analysis suggest
that it
may be possible to use a relatively low energy muon beam that may be available
during the first stages of muon collider construction to probe physics that may
not be accessible elsewhere. We believe that the case for $e\mu$
colliders at high energy is less compelling. It appears that only for some
special flavor
structure of LFV couplings (or when these couplings are $\ll 1$) are $e\mu$
colliders a unique facility for discovering these new interactions.

\acknowledgments
We thank R.~Palmer for communication pertaining to the possibilities for
constructing $\mu e$ colliders. V.~Barger and X.~Tata thank the Aspen Center
for Physics where this work was initiated. This research was supported in part
by the U.S.~Department of Energy under Grants No.~DE-FG03-94ER40833 and
No.~DE-FG02-95ER40896, and in part by the University of Wisconsin Research
Committee with funds granted by the Wisconsin Alumni Research Foundation.


\begin{thebibliography}{99}
\bibitem{note} Equation (\ref{Nf}) also shows that the event rate is not
increased even if the resonance happens to be wide because it can decay
via many different channels, contrary to the statement in Ref.~\cite{choi}.
\bibitem{FN}
Such a neutral particle might also be detectable via a search for
$e^-e^- \to \mu^-\mu^-$ at electron electron colliders, but the
cross section will not be resonance-enhanced.
\end{thebibliography}
\end{document}